\documentclass[10pt,english,twocolumn]{revtex4}
\usepackage[T1]{fontenc}
\usepackage[latin9]{inputenc}
\usepackage{amsmath}
\usepackage{graphicx}
\usepackage{esint}
\makeatletter
\makeatletter
\usepackage{babel}

\makeatother

\usepackage{babel}
\begin{document}

\title{Spontaneous decay in arbitrary cavity size}

\author{G. Flores-Hidalgo}
\email{gfloreshidalgo@unifei.edu.br}
\affiliation{Instituto de F\'{\i}sica e Qu\'{\i}mica, Universidade Federal de Itajub\'a, 37500-903, MG,  Brazil.}

\author{M. M. Silva}
\email{marlon.silva@ufv.br}
\affiliation{Departamento de F\'{\i}sica, Universidade Federal de Vi\c{c}osa, 36570-900, MG, Brazil.}
  
\author{Onofre Rojas} 
\email{ors@dfi.ufla.br}
\affiliation{Departamento de F\'{\i}sica, Universidade Federal de Lavras, 37200-000, MG,  Brazil.}

\begin{abstract}
We consider a complete study of the influence of the cavity size on the spontaneous decay of an atom excited state, roughly  
approximated by a harmonic oscillator. We confine the oscillator-field system in a spherical cavity of radius $R$, perfectly reflective, and work in the formalism of dressed coordinates and states, which allows  to perform non-perturbative calculations for the
probability of the atom to decay spontaneously from the
first excited state to the ground state.
In free space, $R\to\infty$, we obtain known exact results an
for sufficiently small $R$ we have developed a power expansion calculation in this parameter.  Furthermore, for arbitrary cavity size radius, we developed  numerical computations and showed complete agreement with the exact one for $R\to\infty$ and the power expansion results for small cavities,
in this way showing the robustness of our results. We have found that in general the spontaneous decay of an excited state of the atom
increases with the cavity size radius and vice versa. For sufficiently small cavities  the atom practically does not suffers spontaneous decay, 
whereas for large cavities the spontaneous decay approaches the free-space $R\to\infty$ value. On the other hand, for some particular values of the cavity radius, in which the cavity is in resonance with the natural frequency of the atom, the spontaneous decay transition probability  is increased compared to the
free-space case.  Finally, we showed how the probability spontaneous decay  go from an oscillatory time behaviour, for finite cavity radius, to an
almost exponential decay, for free space. 
\\
\\
 PACS numbers: 03.65.Ca, 32.80.Pj
\end{abstract}

\maketitle

\section{Introduction}
The study of spontaneous decay in cavities is an interesting issue, that attracted the interest over the years,  and which was started in Ref. \cite{Purcell},
where it was showed the enhancement of the spontaneous decay for atomic transitions in resonance with the cavity modes.
Many years after, in Ref. \cite{Kleppner},
following the same ideas, it was showed the suppression of the spontaneous decay of an excited atomic state placed in a sufficiently small cavity.
Afterward, many works has been devoted to the subject, both experimental \cite{Gabrielse, hulet,haroche2,David,Goy} and theoretically
\cite{Dowling,Trong,Gieben,Milonni,Milonni2,Stobinska,Barut,Alber}. However to our best knowledge, there has been not considered a full study of the spontaneous decay for arbitrary cavity size. In this paper we consider this task by using a simplified model for the atom, roughly approximated by a one harmonic oscillator and the electromagnetic field is taken as massless scalar field. Obviously, real atoms do not have equally spaced energy levels, 
they are not one dimensional systems. However, our main purpose in this work in not to understand how the nature of the atom affects the spontaneous decay, but about how it depends in a precise way on the cavity size.

To mimic the atom by a harmonic oscillator, with regard to the stability of the ground state, we will use the dressed coordinates and states
approach introduced in Ref. \cite{adolfo1} as a method to account, in a non perturbative way, for
the oscillator radiation process in free space. In subsequent works the concept was used to study the spontaneous emission
of atoms in small cavities \cite{adolfo2}, the quantum Brownian motion \cite{gabriel,adolfoan}, the thermalization process
\cite{gabrielsolo,gabrielyony,thermalst}, the time evolution of bipartite systems \cite{bipartite0,bipartite1,bipartite2}, the entanglement
of biatomic systems \cite{entangled1,entangled2,onofre} and other related issues\cite{nonlinear,yony,eletromag,casana2,casana}.
For a clear explanation see reference \cite{yony}. The formalism
showed to have the technical advantage of allowing exact computation
of the probabilities associated with the different oscillator (atom)
radiation processes \cite{casana}. For example, it has been obtained
easily the probability of the atom to decay spontaneously from the
first excited state to the ground state for arbitrary coupling constant,
weak or strong.
Nevertheless, in all above References exact computations have been possible only for free-space and
for sufficiently small cavity radius it has been possible to do only roughly estimations. The purpose of this work is to present 
techniques applicable to cavities of arbitrary size.
Although the developed computational techniques are applicable to all problems cited above, we will restrict here to compute
the transition probability, due to spontaneous decay,  of the first excited state of an atom, approximated by the dressed harmonic oscillator.

For sufficiently small cavities we will develop a power  expansion computation in the  cavity radius parameter, where
all the coefficients of the expansion are in principle calculable. On the other hand for arbitrary values of the cavity radius we will
consider numerical computations. Finally, we will compare these numerical computations with the one
we developed for small cavities and with the exact result, $R\to\infty$, finding good agreement.

This work is organized as follows. In section II briefly we review the concept of dressed coordinates
and states. In section III we consider the analytical  computation of the transition probability, due to spontaneous decay,  
in the free-space case, $R\to\infty$, and for sufficiently small cavity radius. In section IV we present numerical computations 
for arbitrary cavity size radius and finally in section V we give our concluding remarks. 

\section{Dressed coordinates and states}
The model we consider is a harmonic oscillator linearly interacting with a massless scalar field, the whole system confined inside
a spherical cavity of radius $R$. The Hamiltonian of the system can be put in the form \cite{adolfo1}
\begin{eqnarray}
H=&&\frac{1}{2}\left( p_0^2+\omega_0^2q_0^2\right)+ 
\frac{1}{2}\sum_{k=1}^N\left( p_k^2+\omega_k^2q_k^2\right)\nonumber\\
&& -\sum_{k=1}^Nc_kq_0q_k
+\frac{1}{2}\sum_{k=1}^N\frac{c_k^2}{\omega_k^2}q_0^2,
\label{hamiltonian}
\end{eqnarray}
where $q_0, p_0$ are the oscillator degrees of freedom and $q_k, p_k$ are the corresponding ones for the field modes of
frequencies $\omega_k=\pi k/R$; $k=1,2,3,...;$ $c_k=\eta\omega_k$, $\eta=\sqrt{2g\Delta\omega}$,
$\Delta\omega=\omega_{k+1}-\omega_k=\pi/R$ and $g$ is a frequency dimensional coupling constant. In Eq. (\ref{hamiltonian})
the limit $N\to\infty$ must be taken at the end of calculations and the last term can be seen as a frequency renormalization that
guarantees a positive defined Hamiltonian \cite{weiss}. We diagonalize Hamiltonian (\ref{hamiltonian}) introducing collective coordinates and
momenta $Q_r, P_r$, as
\begin{equation}
q_\mu=\sum_r t_\mu^rQ_r,~~~p_\mu=\sum_r t_\mu^rP_r,~~~\mu=0,k.
\label{bartocolective}
\end{equation}
Substituting above expressions in (\ref{hamiltonian}) we get
\begin{equation}
H=\frac{1}{2}\sum_r\left( P_r^2+\Omega_r Q_r^2 \right),
\label{collectiveh}
\end{equation}
where the matrix elements $t_\mu^r$ are given by
\begin{equation}
t_k^r=\frac{\eta\omega_k}{\omega_k^2-\Omega_r^2}t_0^r,~~~
t_0^r=\left(1+\eta^2\sum_{k=1}^N\frac{\omega_k^2}{(\omega_k^2-\Omega_r^2)^2}\right)^{-\frac{1}{2}}
\label{matrixt}
\end{equation}
and the collective frequencies $\Omega_r$ are solutions of the equation,
\begin{equation}
\omega_0^2-\Omega^2=\eta^2\sum_{k=1}^N\frac{\Omega^2}{\omega_k^2-\Omega^2}.
\label{collectivef}
\end{equation}
Next we introduce dressed coordinates and states as the physically measurable ones, for details see Ref. \cite{yony}.
Denoting the dressed coordinates as
$q_\mu'$, we have that $q_0'$ represents the coordinate of the dressed harmonic oscillator of frequency $\omega_0$ and
the coordinates $q_k'$ represent the coordinates of the dressed field modes of frequencies $\omega_k$. In terms of
the dressed coordinates we introduce the dressed states,
\begin{equation}
\psi_{n_0,n_1,n_2,...}(q')=\prod_{\mu=0}^N\psi_{n_\mu}(q_\mu'),
\label{dresseds}
\end{equation}
where $\psi_{n_\mu}(q_\mu')$ is eingenfunction of one dimensional harmonic oscillator of frequency $\omega_\mu$ and energy
$(n_\mu+1/2)\omega_\mu$. Physically, the dressed state (\ref{dresseds}) is the one in which the dressed harmonic oscillator
is its $n_0$-th energy level and the field system is a state in which there are $n_k$ field quanta of frequencies $\omega_k$. Requiring the
stability of the dressed oscillator ground state  in the absence of field quanta, we get  the relation between dressed
coordinates and the collective ones. To assure the stability of the state $\psi_{0,0,0,...}(q')$ we identify it with the ground state
of the total system Hamiltonian (\ref{collectiveh}), from which we get, 
\begin{equation}
q_\mu'=\sum_r \sqrt{\frac{\Omega_r}{\omega_\mu}} t_\mu^r Q_r\,.
\label{dressedtocolective}
\end{equation}
The relation between dressed and collective coordinates allows to do computations of transitions amplitudes between the
dressed states (\ref{dresseds}). If at $t=0$, the oscillator-field system is in the state $|n_0,n_1,n_2,...\rangle_d$ (whose coordinate 
representations is given by $\psi_{n_0,n_1,n_2,...}(q') $), the probability amplitude to find the system at $t>0$ in the state
$|m_0,m_1,m_2,...\rangle_d$ is given by
\begin{equation}
{\cal A}_{n_0,n_1,n_2,...}^{m_0,m_1,m_2,...}(t)=~_d~\!\!\langle m_0,m_1,m_2,...|{\rm e}^{-i\hat{H}t}
|n_0,n_1,n_2,...\rangle_d
\label{amplitude}
\end{equation}
As showed in Ref. \cite{casana}, the different transition amplitudes above can be cast in terms of
\begin{equation}
f_{\mu\nu}(t)=\sum_{r} t_\mu^r t_\nu^r {\rm e}^{-i\Omega_r t}.
\label{fmunu}
\end{equation}
For  example, considering as initial state, the one in which the dressed harmonic oscillator is in its first excited level and there are no field
quanta, we get for the probability amplitude for the system to remain in the same initial state,
\begin{equation}
{\cal A}_{1,0,0,0...}^{1,0,0,0,...}(t)=f_{00}(t).
\label{ampltude}
\end{equation}
Also,  the probability amplitude for the dressed harmonic oscillator to decay spontaneously from their first excited
level to the ground state, by emission of a field quanta of frequency $\omega_k$, is given by
\begin{equation}
{\cal A}_{1,0,0,0,...}^{0,0,...,0,1_k,0,0...}=f_{0k}(t).
\label{decay}
\end{equation}
In any case, the computation of $f_{00}(t)$ or $f_{0k}(t)$ is a formidable task: we have to sum infinite terms that we don't know
a priori. This, because each term of the sum depends on $\Omega_r$, given as the solutions of Eq. (\ref{collectivef}), which in the limit
$N\to\infty$,  possess infinite solutions that can not be computed analytically. Nevertheless, in the limit in which
the cavity radius $R\to\infty$,  it is possible to get closed expressions for $f_{\mu\nu}(t)$ \cite{gabrielsolo}. In next section
we extend the method for arbitrary cavity radius and perform analytical calculations for sufficiently small $R$.

\section{Analytical Computation of probability amplitudes}
Following Ref. \cite{gabrielsolo}
we introduce a complex variable function,
\begin{equation}
W(z)=z^2-\omega_0^2+\sum_{k=1}^N\frac{\eta^2 z^2}{\omega_k^2-z^2}.
\label{function}
\end{equation}
Note that the real roots of $W(z)=0$ are precisely the solutions of Eq. (\ref{collectivef}). Also, from Eq. (\ref{matrixt}) we can
write for $t_0^r$,
\begin{equation}
(t_0^r)^2=\frac{2\Omega_r}{W'(\Omega_r)}
\label{t0r2}
\end{equation}
where $W'(z)=dW/dz$. Taking $\mu=\nu=0$  in Eq. (\ref{fmunu}) and using (\ref{t0r2}) we get
\begin{eqnarray}
f_{00}(t)&=&\sum_{r}\frac{2\Omega_r{\rm e}^{-i\Omega_r t}}{W'(\Omega_r)}\nonumber\\
&=&\frac{1}{\pi i}\oint_C\frac{z{\rm e}^{-izt}}{W(z)} dz
\label{f00}
\end{eqnarray}
where in passing to the second line we have used residue theorem and $C$ is a counter-clockwise contour that encircles the real
positive roots $\Omega_r$, {\it i.e}, a contour that encircles the real positive axis in the complex plane $z$.
In the same way we obtain for (\ref{decay})
\begin{equation}
f_{0k}(t)=\frac{\eta\omega_k}{\pi i}\oint_C\frac{ze^{-izt}}{W(z)(\omega_k^2-z^2)}dz.
\label{fok}
\end{equation}
Using  $\omega_k=k\pi/R$ and identity,
\begin{eqnarray}
\sum_{k=1}^{N\to\infty}\frac{\alpha^2}{k^2-\alpha^2}=\frac{1}{2}(1-\pi\alpha\cot(\pi \alpha)).
\label{formula}
\end{eqnarray}
we get for  $W(z)$,
\begin{eqnarray}
\label{pc01}
W(z)=z^2-\omega_0^2+\frac{g\pi}{R}[1-zR\cot(zR)].
\end{eqnarray}
 Formulas (\ref{f00}) and (\ref{fok}) are valid for
any arbitrary cavity radius $R$, however we can go further analytically only in the two extreme cases: very large and
sufficiently small cavity radius.
For very large cavity, the free space,  considering $z=x\pm i\epsilon$ and $\epsilon\to 0^+$, we get for
(\ref{pc01}) in the limit $R\to\infty$
\begin{equation} 
W(x\pm i\epsilon) = x^2-\omega_0^2\mp ig\pi.
\label{wrtoinfty}
\end{equation}
Using above expression in (\ref{f00}) and considering a rectangular contour for $C$,  we get for the probability amplitude 
of the harmonic oscillator to remain in the first excited state, in free space \cite{gabrielsolo}
\begin{equation}
f_{00}(t)=2g\int_{0}^\infty dx\frac{x^2 e^{-ixt}}{(x^2-\omega_0^2)^2+\pi^2g^2x^2},
\label{Rtoinfty}
\end{equation}
from which we have $f_{00}(t\to\infty)=0$. For finite values of $t$ we can perform easily a numerical computation of above
integral and we get an almost exponential decay function for the corresponding probability \cite{gabrielsolo}. 

\subsection{Computation of $ f_{00}(t)$ for small $R$}

For arbitrary cavity radius it is not possible to get analogous results as (\ref{Rtoinfty})  and we limit here to the case in which
the cavity radius is sufficiently small. In this case, using (\ref{pc01}), it is convenient to write  Eqs. (\ref{f00}) and (\ref{fok}) as,
\begin{eqnarray}
\label{rp03}
f_{00}(t)=\frac{1}{\pi i}\oint_C \frac{z e^{-izt/R}}{z^2-R^2\omega_0^2+\pi R g[1-z\cot(z)]}dz,
\end{eqnarray}
\begin{equation}
f_{0k}(t)=\frac{\eta\omega_k R^2}{\pi i}\oint_C
 \frac{z(R^2\omega_k^2-z^2)^{-1} e^{-izt/R}}{(z^2-R^2\omega_0^2+\pi R g[1-z\cot(z)])}dz.
\label{fok1}
\end{equation}
where we made the change of variable $zR\to z$  in both integrals above.  
Since $Rg$ is an dimensionless quantity, expanding the denominator of (\ref{rp03}) in powers of $\pi Rg$, we have
\begin{eqnarray}
\label{pc022}
f_ {00}(t)=&& \sum_{j=0}^\infty (\pi R g)^j a_j,
\label{f00exp}
\end{eqnarray}
where 
\begin{equation}
a_j=\frac{(-1)^j}{\pi i}\oint_C \frac { z(1 - z\cot (z))^je^{-itz/R}} {(z^2 -
     R^2\omega_ 0^2)^{j+1}} dz.
\label{coefficient}
\end{equation}
Using residue theorem, we can compute all above coefficients. We have for the first term
\begin{eqnarray}
a_0&=&\frac{1}{\pi i}\oint \frac { ze^{-itz/R}} {(z^2 -
     R^2\omega_ 0^2)} dz\nonumber\\
&=&e^{-i\omega_0 t}
\label{a0}
\end{eqnarray}
where we used the fact that the pole $z=R\omega_0$ is the only one inside $C$. Taking $j=1$ in Eq. (\ref{coefficient}) we have
for $a_1$,
\begin{eqnarray}
a_1=-\frac{1}{\pi i}\oint_C \frac { z(1 - z\cot (z))e^{-itz/R}} 
{(z^2 -     R^2\omega_ 0^2)^{2}} dz.
\end{eqnarray}
In this case we have the second  order  pole $R\omega_0$ and first order poles $0,~\pi,~2\pi,~3\pi,...$. Using residue theorem
we have
\begin{eqnarray}
a_1=&&-\frac{e^{-i \omega_0 t}}{2 R^2\omega_0^2} \bigg\{-i \omega_0 t+ (i\omega_0 t-1) R\omega_0 \cot(R \omega_0)\nonumber\\
&&+\frac{R^2\omega_0^2}{\sin^2(R \omega_0)}\bigg\}
+\sum_{n=1}^{\infty}\frac{2 e^{-i\omega_n t} \omega_n^2}{R^2(\omega_n^2 -\omega_0^2)^2}.
\label{a1}
\end{eqnarray}
All other terms (\ref{coefficient}) can be calculated noting that $R\omega_0$ is a pole of order $j+1$ whereas the other poles,
$0,~\pi, ~ 2\pi,~3\pi$,... are of order $j$. Although all the coefficients $a_j$ are computable,  for higher
$j$ the expressions are cumbersome. Here we quote only $a_2$,  given by
\begin{eqnarray}
a_2=&&\frac{e^{-i \omega_0 t}}{8 (R\omega_0)^4}\bigg\{-\omega_0^2 t^2-i \omega_0 t (-1+4 R^2\omega_0^2)\nonumber\\
&&+2 \big(R^2\omega_0^2+R^4\omega_0^4\big)
+2 R\omega_0 \big(1+\omega_0^2 t^2\nonumber \\
&&-5 R^2\omega_0^2
+i \omega_0 t \left(1+2 R^2\omega_0^2\right)\big) \cot(R \omega_0)\nonumber\\
&&+R^2\omega_0^2 \big(2
-7 i \omega_0 t-\omega_0^2 t^2+8 R^2\omega_0^2\big) \cot^2(R \omega_0)\nonumber\\
&& +2 i (5 i+2 \omega_0 t) R^3\omega_0^3 \cot^3(R \omega_0)\nonumber\\
&&+6 R^4\omega_0^4 \cot^4(R \omega_0)\bigg\}\cr
& &-\sum_{n=1}^{\infty}\frac{2 i e^{-i\omega_n t} \omega_n^2}
{R^4\omega_0 \left(\omega_n^2-\omega_0^2\right)^4}
\big(\omega_n^3 \omega_0 t-5 i \omega_n^2 \omega_0\nonumber\\
&&-\omega_n  \omega_0^3 t -i \omega_0^3\big).
\label{a2}
\end{eqnarray}
%
%
\begin{figure}[b!]
\includegraphics[scale=0.5]{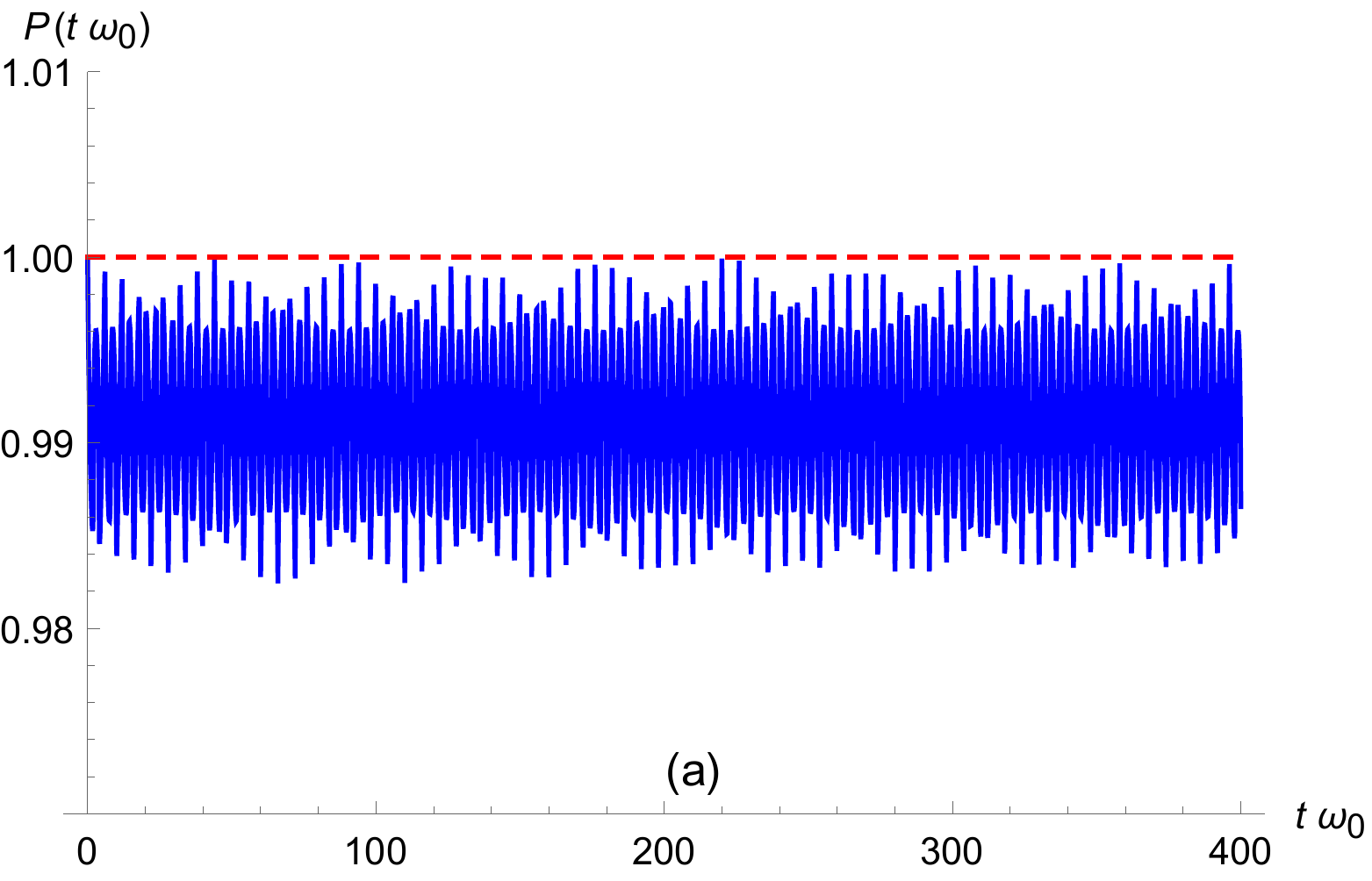}\\
\includegraphics[scale=0.5]{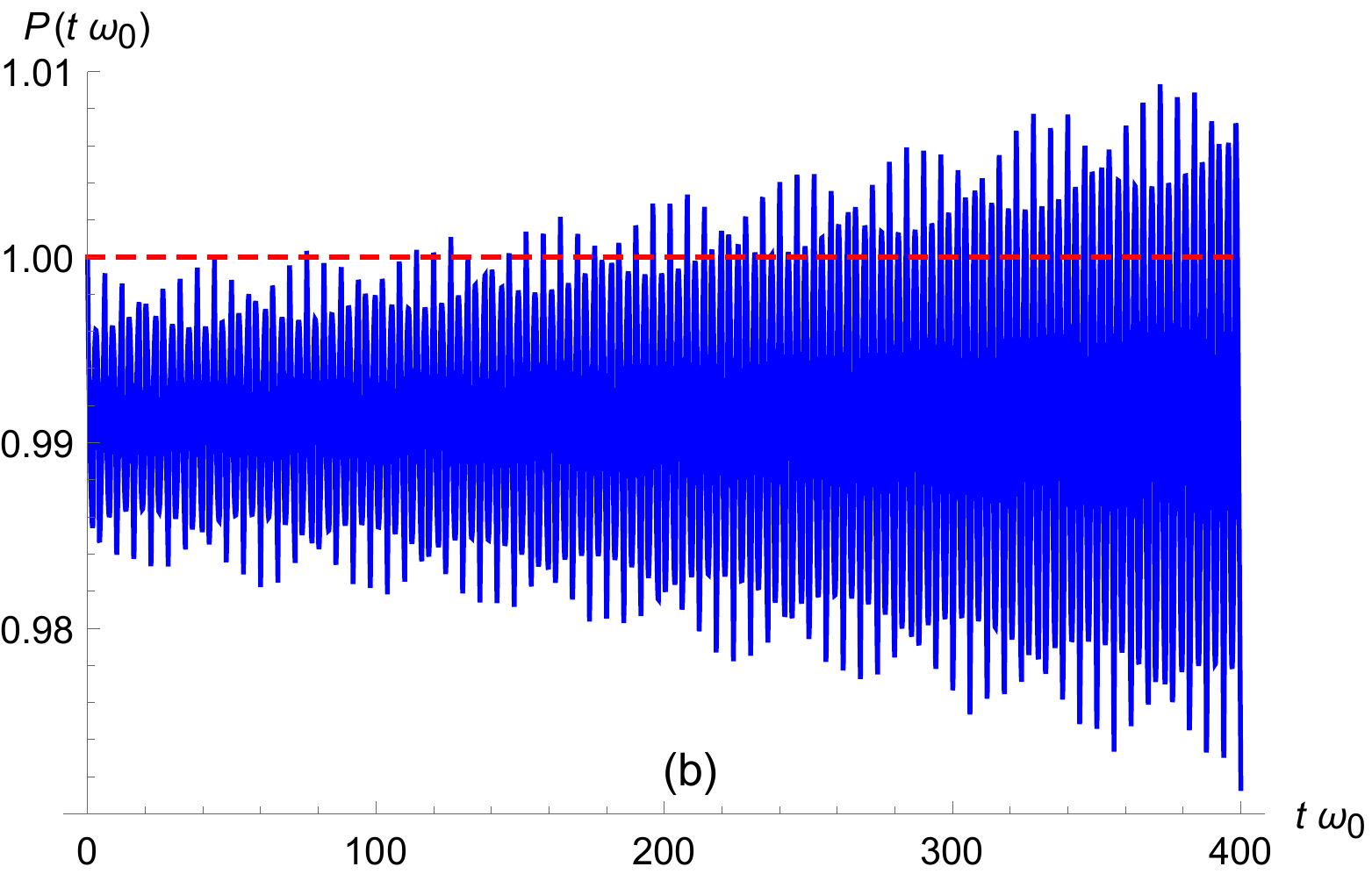}\\
\includegraphics[scale=0.5]{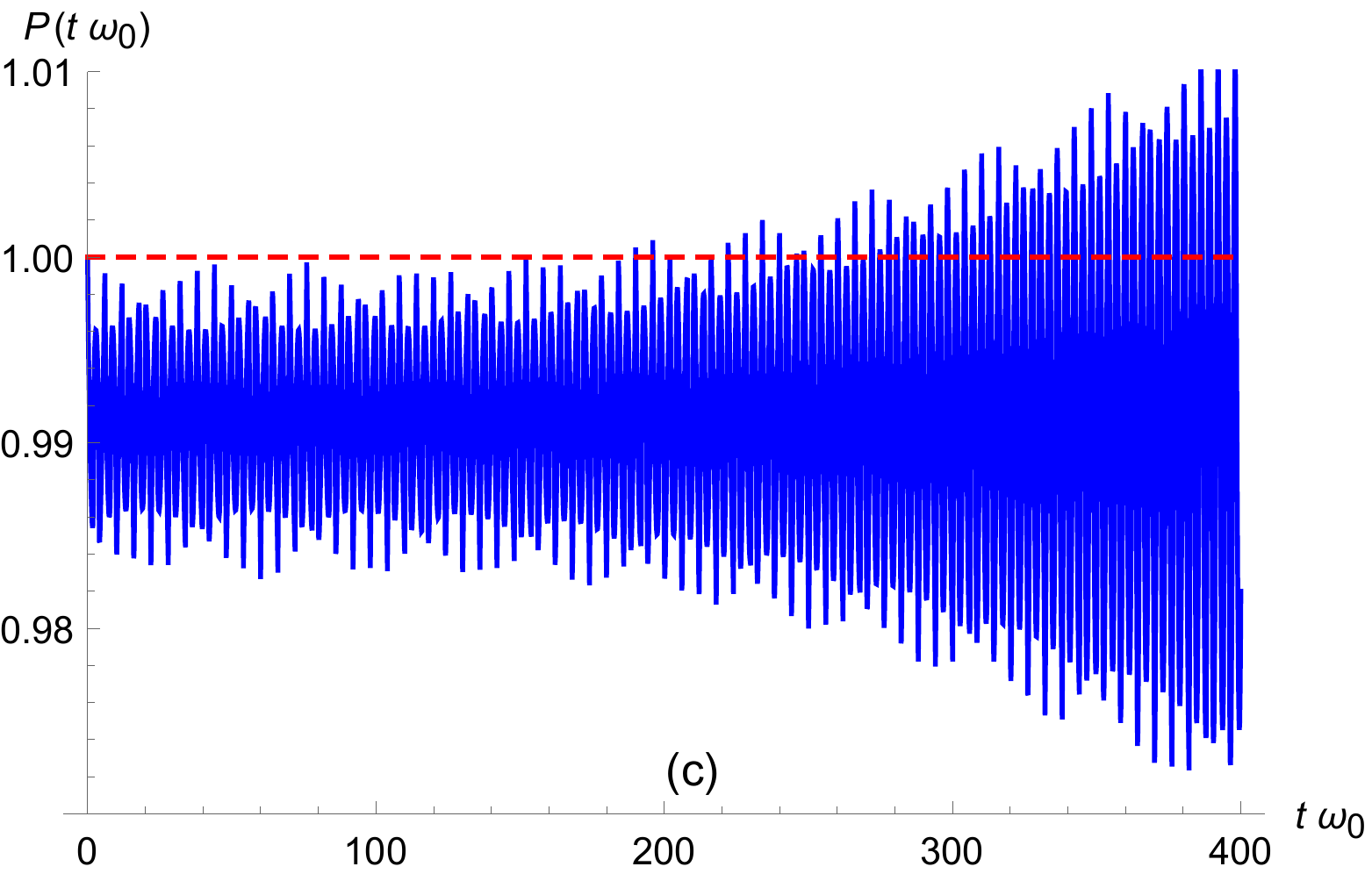}\\
\includegraphics[scale=0.5]{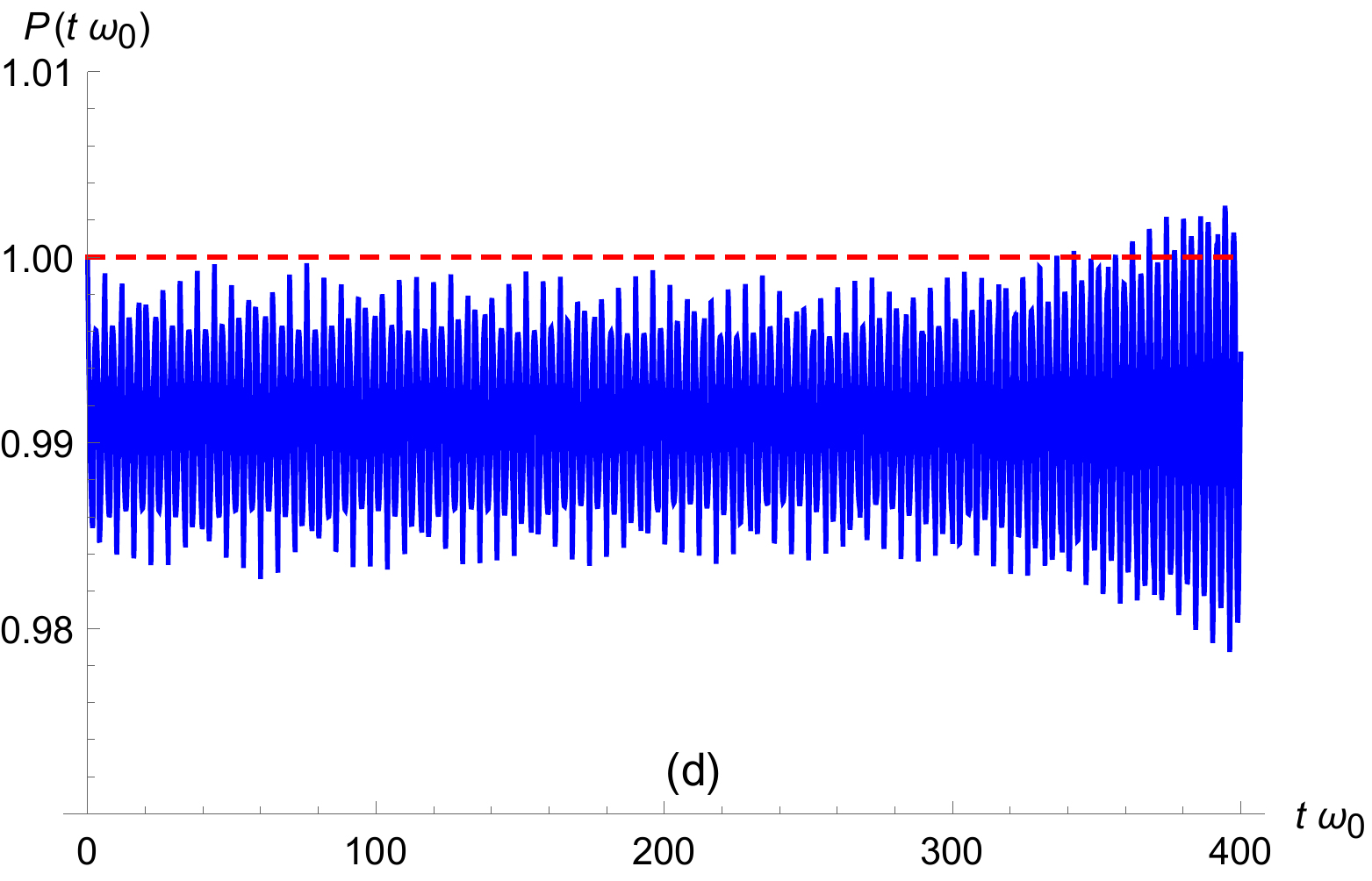}
 \caption{(Color online)  $\omega_0R=1$, $g=\omega_0/274$.}
\label{fig1} 
\end{figure}
%
We have to remark that  the infinite sums in expressions (\ref{a1}) and (\ref{a2}) are  fastly convergent.  From Eq. (\ref{f00exp}) we have
for $|f_{00}(t)|^2$,
\begin{equation}
|f_{00}(t)|^2=\sum_{j,l=0}^\infty (\pi R g)^{j+l} a_j a_l^\ast,
\label{|foo|}
\end{equation}
which at second order  in $(\pi Rg)$ is 
\begin{eqnarray}
|f_{00}(t)|^2=&&1+\pi Rg\left(e^{-i\omega_0 t}a_1^\ast+e^{i\omega_0 t}a_1\right)\nonumber\\
&&
+
(\pi Rg)^2\left(e^{-i\omega_0 t}a_2^\ast+e^{i\omega_0 t}a_2+|a_1|^2\right)\nonumber\\
&&+{\cal O}\left((\pi R g)^3\right).
\label{foot3}
\end{eqnarray}
We stress here that $|f_{00}(t)|^2$, given by (\ref{|foo|}), satisfies at each order in $gR$, $|f_{00}(0)|^2=1$. 
Before to consider  numerical computations for (\ref{|foo|}), we discuss the question about the validity of
the expansion (\ref{f00exp}). This series expansion will be convergent if
\begin{equation}
\lim_{j\to\infty} \frac{(\pi R g)^{j+1}|a_{j+1}|}{(\pi R g)^j|a_j|}<1.
\end{equation}
Since in the limit $j\to\infty$ the order of the poles in $a_j$ and $a_{j+1}$ are almost the same, we get $\lim_{j\to\infty}|a_{j+1}|/|a_j|=1$,
and  the condition for convergence of  $(\ref{f00exp})$ is given by
\begin{equation}
\pi R g<1.
\label{validity}
\end{equation}
Consequently, for fixed parameter $g$, the cavity radius $R$ must be considered  small if $R<1/(\pi g)$, a condition
independent of time and of the frequency oscillator $\omega_0$. On the other hand if we consider very small cavities, 
$R<<1/(\pi g)$, we expect that the first terms in the power expansion (\ref{f00exp}) to give the relevant contributions,
at least for time values not sufficiently large. This last condition can be inferred from expressions for $a_1$ or $a_2$, for
example, where we can note that such coefficients increases with time almost linearly or quadratically. Therefore, if the
series expansion is truncated, the corresponding probability (\ref{|foo|}), could violate
the upper limit $|f_{00}(t)|^2\leq 1$ for time values sufficiently large. For such time values, we have to consider higher order
contributions. As illustration, we consider a very small cavity with  $g=\omega_0/274$ and $\omega_0R=1$, where 
$R=1/(274 g)<<1/(\pi g)$. 
In Fig. \ref{fig1} we depict $P(t)=|f_{00}(t)|^2$, as given by (\ref{|foo|}), with $a_{l}$ computed at first, second, thirth
and fourth order in $(Rg)$.
At  first order  $P(t)\leq 1$,  as expected,
at second order it violates this condition around $\omega_0t \approx80$, at third order $P(t)>1$ around 
$\omega_0t\approx 200$ and at fourth order $P(t)>1$ for $\omega_0 t\approx 350$.  However, if we include higher terms
the behaviour of $P(t)$  improves for larger values of  $\omega_0 t$. For example, at sixth order we display $P(t)$  in
Fig. \ref{fig2},  where we see that the result is valid up to $\omega_0 t=400$. 
\begin{figure}[ht]
\includegraphics[scale=0.55]{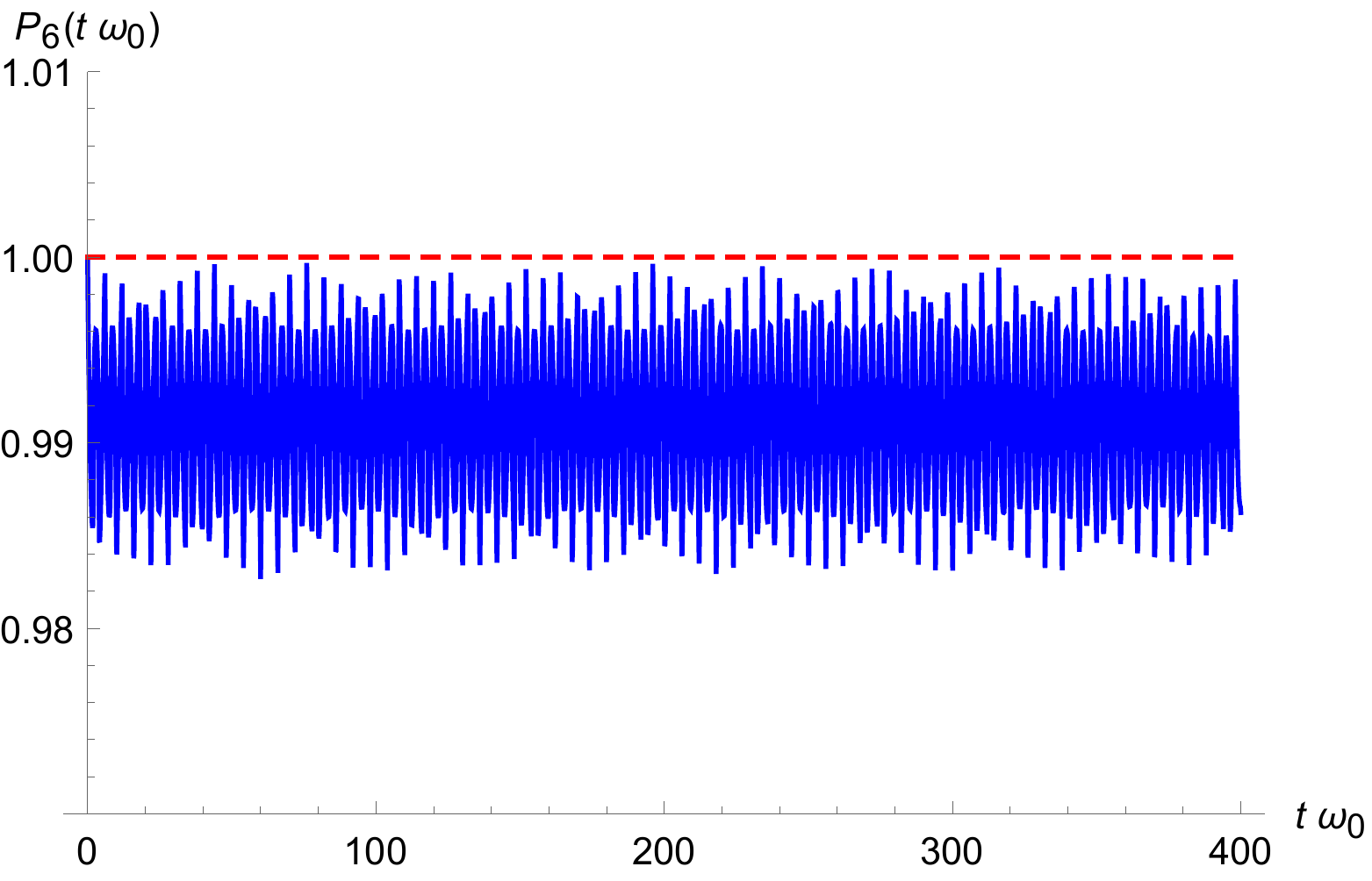}
 \caption{(Color online) $\omega_0R=1$, $g=\omega_0/274$.}
\label{fig2} 
\end{figure}
\begin{figure}[b!]
\includegraphics[scale=0.5]{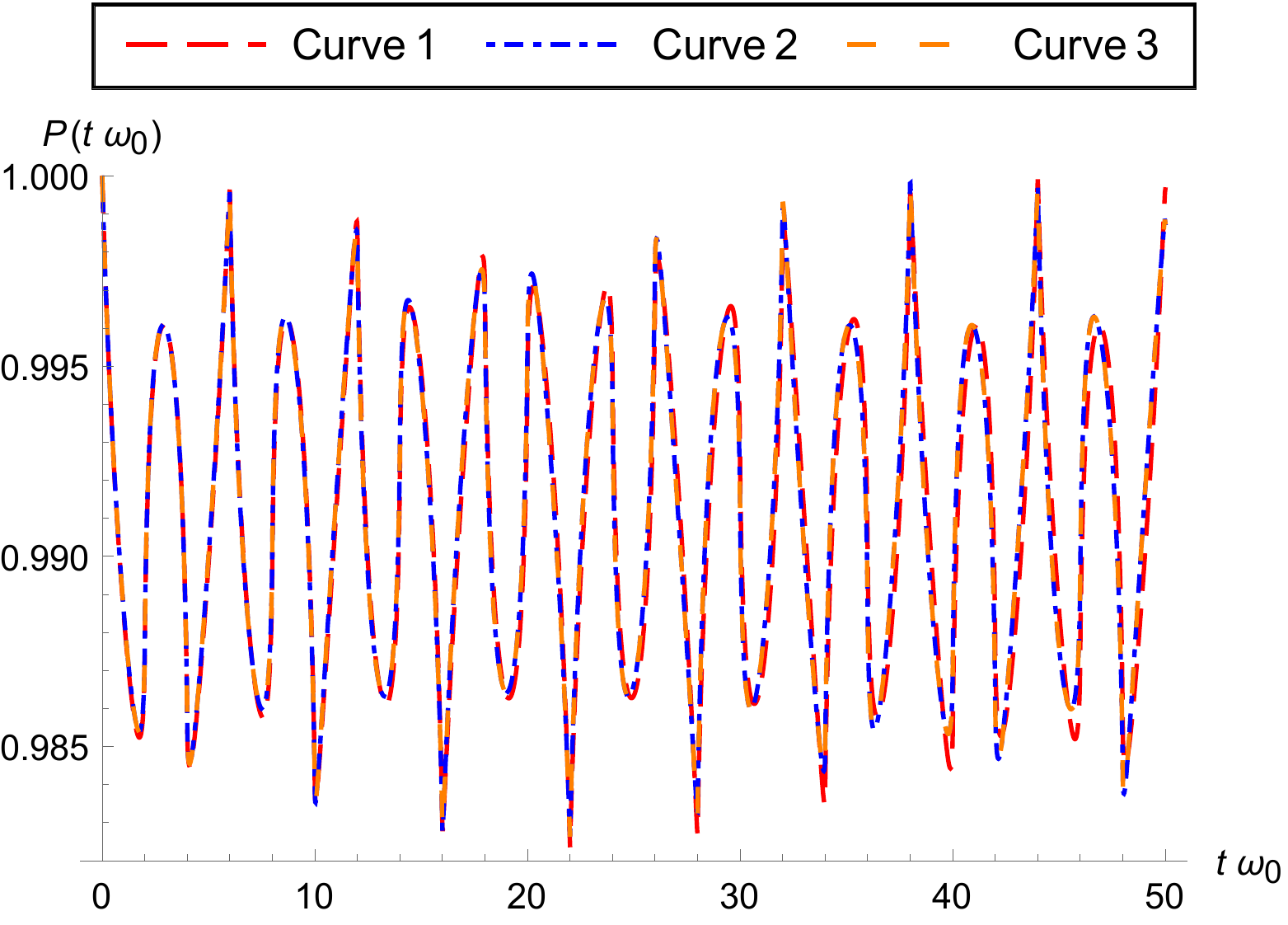}
 \caption{(Color online) Expansion at: Curve 1: First order, Curve 2: Third order, Curve 3: Sixth order; $\omega_0R=1$, $g=\omega_0/274$.}
\label{fig3} 
\end{figure} 
On the other hand if we compare
$P(t)$ at different orders of approximations, we find that for $\omega_0 t$ not sufficiently large, all approximations
gives almost the same results, as one can conclude from Fig. \ref{fig3}, where we compare $P(t)$ at first, third, and
sixth order. Note, that very small differences only appears for sufficiently large values of $\omega_0 t$ and if this parameter
is not large enough, the results are almost the same. If we consider other values for the cavity radius in the regime $R<<1/(\pi g)$,
we get almost the same results above.
Therefore, we conclude that for a sufficiently small cavity
higher order corrections terms in the expansion (\ref{|foo|}) will be important only for large values of $\omega_0 t$. 

But what, about the physical meaning of our results? From
Fig. \ref{fig2}, we see that $P(t)$ oscillates around $0,991$, never decreasing than $0,982$ and we can conclude 
 that for the very small cavity radius considered, we have that the probability of the dressed harmonic oscillator 
to remain in their first excited level is around $99,1$ \%. We have the inhibition  of the spontaneous decay similar to the pointed
out for the first time in Ref. \cite{Kleppner}. If we consider other values for the cavity radius less than the one considered above,
the probability $P(t)$ increases, that is, the spontaneous decay of the first excited state is more and more suppressed as $R$ decreases.  
In order to appreciate the orders of magnitude involved in this phenomenon, we consider SI units, in this case $\omega_0  R=1$
can be replaced by $\omega_0 R/c=1$ from which considering $\omega_0\sim4\times 10^{14}$ s, in the visible, red we get
$R\sim 7.5 \times 10^{-7}$ m and for $\omega_0\sim 2\times 10^{10}$ s, a typical microwave frequency, we have 
$R\sim 1.5\times 10^{-2}$ m. For these parameter values we expect an almost stability of atomic excited levels.

\subsection{Computation of $f_{0k}(t)$}
As done for $f_{00}(t)$, expanding the denominator of Eq. (\ref{fok1}) in powers of $Rg$, we get,
\begin{equation}
f_ {0k}(t)=\eta \omega_k \sum_{j=0}^\infty (\pi Rg)^jb_j,
\label{fok2}
\end{equation}
where
\begin{equation}
b_j=
\frac{(-1)^j R^2}{\pi i } \oint\bigg[\frac {z e^{-itz/R} }{(R^2\omega_k^2-z^2)}  \frac { (1 - z\cot (z))^j} {(z^2 -
     R^2\omega_ 0^2)^{j+1}}\bigg] dz.
\label{bj}
\end{equation}
All above coefficients can be computed using residue theorem, where the pole $R\omega_k=\pi k$ is of order $(j+1)$,
the pole $R\omega_0$ is of order $(j+1)$ and the poles $0,~\pi,~2\pi,...$ are of order $j$. Because the final expressions
are complicated, we quote only expressions for $b_0$ and $b_1$, respectively given by
\begin{equation}
b_0=\frac{\left(e^{-i\omega_0 t}-e^{-i\omega_k t}\right)}
{\left(\omega_k^2-\omega_0^2\right)},
\label{b0}
\end{equation}
\begin{eqnarray}
b_1=&-&
\frac{e^{-i \omega_0 t }}{2\omega_0^2(\omega_k^2-\omega_0^2)^2}\bigg\{ R^2(-\omega_k^2\omega_0^2+ \omega_0^4) \cot^2(R\omega_0)
\nonumber\\
& +&i \omega_k^2 (\omega_0 t + i R^2\omega_0^2) 
+ \omega_0^2 (-2 - i \omega_0 t + R^2\omega_0^2)\cr
&+&R\omega_0 (\omega_k^2(1 - i \omega_0 t) + (1 + i \omega_0 t) \omega_0^2) \cot(R\omega_0)\bigg\}\cr
&+&\frac{e^{-i\omega_k t } (2 i \omega_k^3  t+7 \omega_k^2  -2 i \omega_k  \omega_0^2 t
+ \omega_0^2)}{2R^2(\omega_k^2-\omega_0^2)^3}\nonumber\\
&+&\sum_{n\neq k}^\infty\frac{2e^{-i\omega_n}\omega_n^2 }{R^2(\omega_k^2-\omega_n^2)
(\omega_n^2-\omega_0^2)^2}.
\label{b1}
\end{eqnarray}
From Eq. (\ref{fok2}) we get for  $|f_{0k}(t)|^2$
\begin{equation}
|f_{0k}(t)|^2=\eta^2\omega_k^2 \sum_{j,l=0}^\infty (\pi Rg)^{j+l} b_jb_l^\ast.
\label{|fok|}
\end{equation}
Although we can perform numerical computations with the obtained expression for $|f_{0k}(t)|^2$ at the
order we desire,  as done for $P(t)$, we have to note that such quantity is in general very small, for sufficiently small cavities, as one can
easily verify from the identity
\begin{equation}
\sum_{k}|f_{0k}(t)|^2+|f_{00}(t)|^2=1,
\label{ident}
\end{equation} 
from which we find $\sum_{k} |f_{0k}(t)|^2=1-P(t)$. For the values just considered above, $\omega_{0} R=1$, $g=\omega_0/274$,
we obtain $\sum|f_{0k}(t)|2<0,018$, that is, the probability for the harmonic oscillator to decay from its first excited state
to the ground state by emission of an arbitrary field quanta is smaller that $1,8$ \%. 

From expressions (\ref{b0}) or (\ref{b1}) it is possible to see that the maximum contribution for $|f_{0k}(t)|^2$, is given by
those values of $\omega_k$ around $\omega_0$. In general for sufficiently small cavity radius, $\omega_k=k\pi/R>\omega_0$ and 
there is no value for $\omega_k$ close enough to $\omega_0$. Consequently,  $|f_{0k}(t)|^2$ will be very small. In other words,
when the cavity size is sufficiently small, there is no field quanta with energy near the gap energy between the first excited energy
level and the ground state, and in this way, the spontaneous decay of the first excited level is practically suppressed. On the other
hand, if we consider cavities where 
$\omega_0=\omega_k$, that is, resonant values of $R=k\pi/\omega_0$, $k=1,\ 2,\ \ldots$ we expect from (\ref{b0})-(\ref{b1}), that
 $|f_{0k}(t)|$ increases appreciably in
relation to the non resonant values. In this case, rigorously, expressions for (\ref{a1}), (\ref{a2}),  (\ref{b0}) and (\ref{b1}) are not valid since
for resonant values of the cavity radius, the poles of $a_j$ and $b_j$ are of order different from the ones considered previously.
Although we have computed the corresponding expressions, we do not present them here, since in general the first terms
of the series expansion for $f_{00}(t)$ or $f_{0k}(t)$, are valid only for initial time values, that is, resonant values of $R$ are small but not not
sufficiently small.
Instead, in next chapter we perform
numerical computations,  for arbitrary cavity radius, where we will show the  enhancement of the spontaneous decay
in resonant cavities whenever  $R=n\pi/\omega_0$, $n=1,\ 2,\ \ldots$.

\begin{figure}[b!]
\includegraphics[scale=0.5]{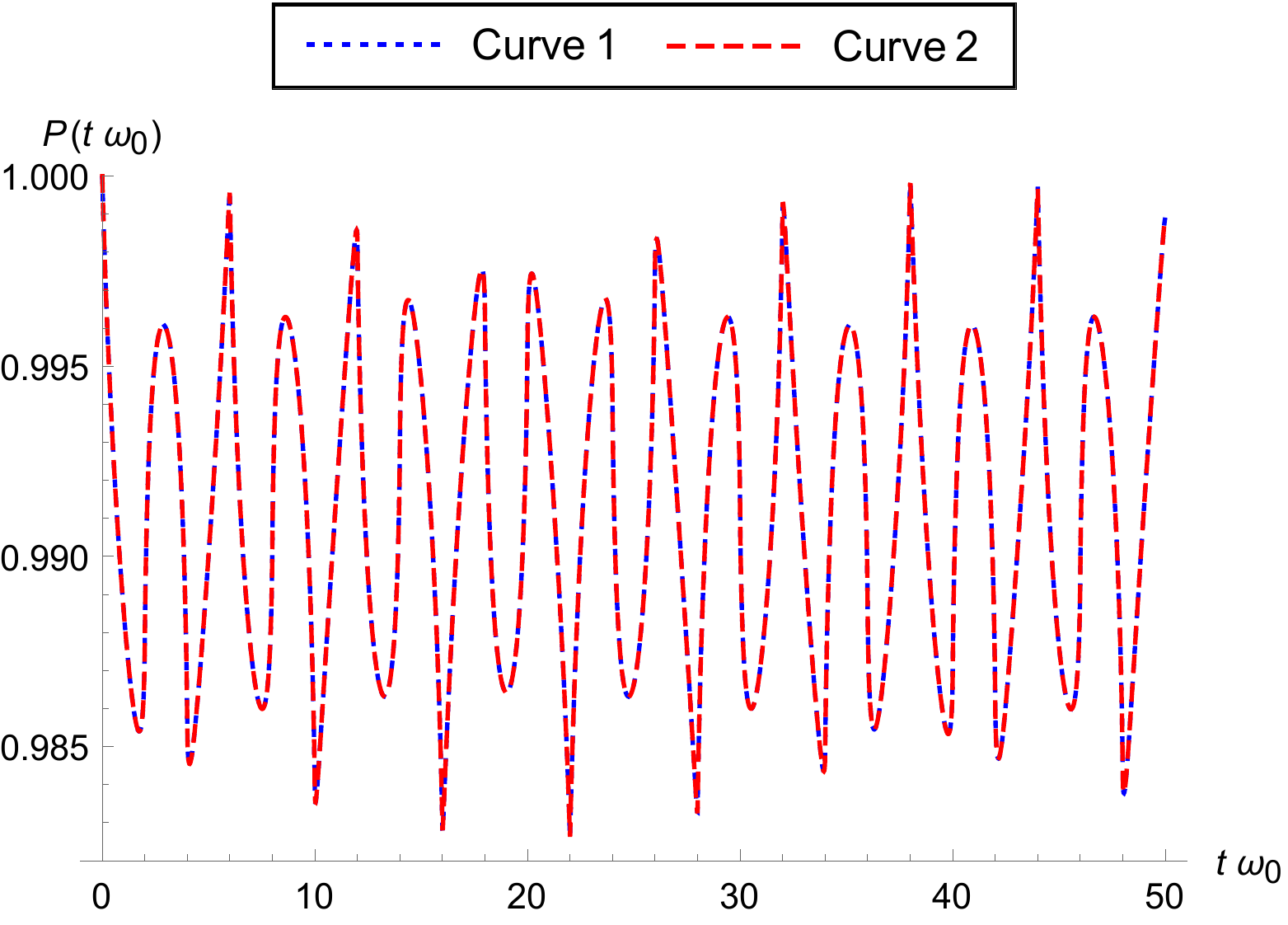}
 \caption{(Color online) Curve 1: Expansion at sixth order;
 Curve 2: Numerical computation.
 $\omega_0R=1$, $g=\omega_0/274$. }
\label{fig4}
\end{figure} 
\section{Arbitrary cavity size: Numerical computations}
We can compute $f_{00}(t)$ or $f_{0k}(t)$ numerically, for arbitrary cavity radius in two ways. First, we can use
expressions (\ref{f00}) or (\ref{fok})  with an appropriate contour $C$ to perform the integral lines
numerically. We can consider for example a rectangular closed contour, with parameters in such a way that this
contour encircles the poles in the real positive axis. This is not an easy task, since given a contour there is no way
to prove that inside the contour  the only poles are those in the real positive axis. Therefore, we have to  proceed iteratively
decreasing the size of the contour in each step until the results stabilizes. However, this becomes in long time computations. 
\begin{figure}[b!]
\includegraphics[scale=0.5]{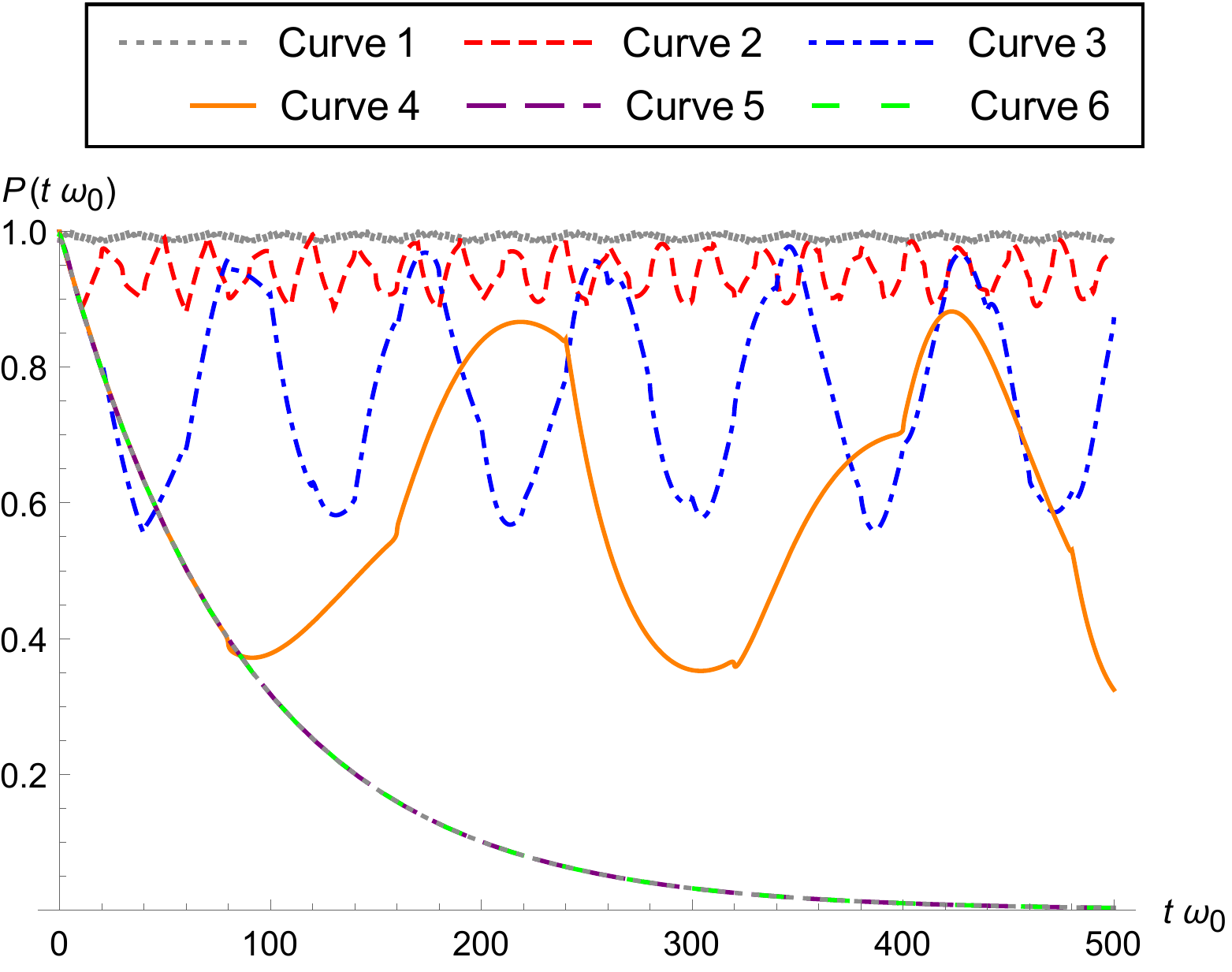}
 \caption{(Color online) Curve 1: $\omega_0R=1$, $g=\omega_0/274$; Curve 2: $\omega_0R=5$, $g=\omega_0/274$; Curve 3: $\omega_0R=10$, $g=\omega_0/274$;
 Curve 4: $\omega_0R=40$, $g=\omega_0/274$; 
 Curve 5: $\omega_0R=400$, $g=\omega_0/274$; Curve 6: $R\rightarrow\infty$, $g=\omega_0/274$.  }
\label{fig5} 
\end{figure}
Another way to compute $f_{00}(t)$ or $f_{0k}(t)$ is to solve for the collective
frequencies $\Omega_r$ from (\ref{collectivef}) numerically and performing the sums in (\ref{fmunu}). But since
it is not possible to solve numerically for all the collective frequencies, we compute only o finite number of
them, for example the first $10^4$ solutions. For the other collective frequencies we can use with good precision $\Omega_k=\omega_k$,
since as $\Omega_r$ increases it approaches $\omega_k$ \cite{adolfo1}.
Also the summation in (\ref{fmunu}) must stop at the maximum values obtained for $\Omega_r$. Again, this could be a problem, 
but as we will show bellow the
sums in (\ref{fmunu}) converges rapidly. Therefore, we will do numerical computations in the way just
described. For this end, first we perform the sums in Eq. (\ref{matrixt}) and (\ref{collectivef}), using (\ref{formula}) we
have respectively
\begin{equation}
(t_0^r)^2=\frac{\eta^2\Omega_r^2}{(\Omega_r^2-\omega_0^2)^2+\frac{\eta^2}{2}(3\Omega_r^2-\omega_0^2)+
\pi^2 g^2\Omega_r^2}
\label{num1}
\end{equation}
and
\begin{equation}
\cot(R\Omega_r)=\frac{\Omega_r}{\pi g}+\frac{1}{R\Omega_r}\left(1-\frac{R\omega_0^2}{\pi g}\right).
\label{num2}
\end{equation}
From last expression we note that $(t_0^r)^2\sim \Omega_r^{-2}$ for large $\Omega_r$, therefore we can compute
\begin{equation}
f_{00}(t)=\sum_{r}(t_0^r)^2 e^{-i\Omega_r t}
\label{num3}
\end{equation}
numerically with a finite number of solutions for $\Omega_r$, large solutions $\Omega_r$ will give negligible
contributions. As a first application, we consider, $\omega_0 R=1$, $g=\omega_0/274$, the case treated in above section.
In this case we get for $P(t)$, the result depicted in Fig. \ref{fig4} as doted line, where 
for comparison,  we plotted the one obtained in the last section as dashed line. We can note that both results are in good agreement.
Next we consider the time behaviour of $P(t)$ for other values of the cavity radius.    In order to compare, the behavior of $P(t)$ for 
increasing values of $R$ we consider $g=\omega_0/274$ fixed and
$Rg=1/274, 5/274,10/274,40/274$, and $ 400/274$. The results for $P(t)$  are depicted in Fig. \ref{fig5}, in the time interval 
$0\leq \omega_0t\leq 500$ . We conclude that $P(t)$, in general, decreases as $R$ increases and vice-versa.
Note that as $R$ increases $P(t)$ approaches the free-space case, $R\to\infty$, whereas for very small cavities the probability of spontaneous
decay practically go to zero, $P(t)\approx 1$. For $R$ finite, $P(t)$  is an  almost oscillating  function of $\omega_0 t$, whose  period increases
with $R$. 
\begin{figure}[b!]
\includegraphics[scale=0.5]{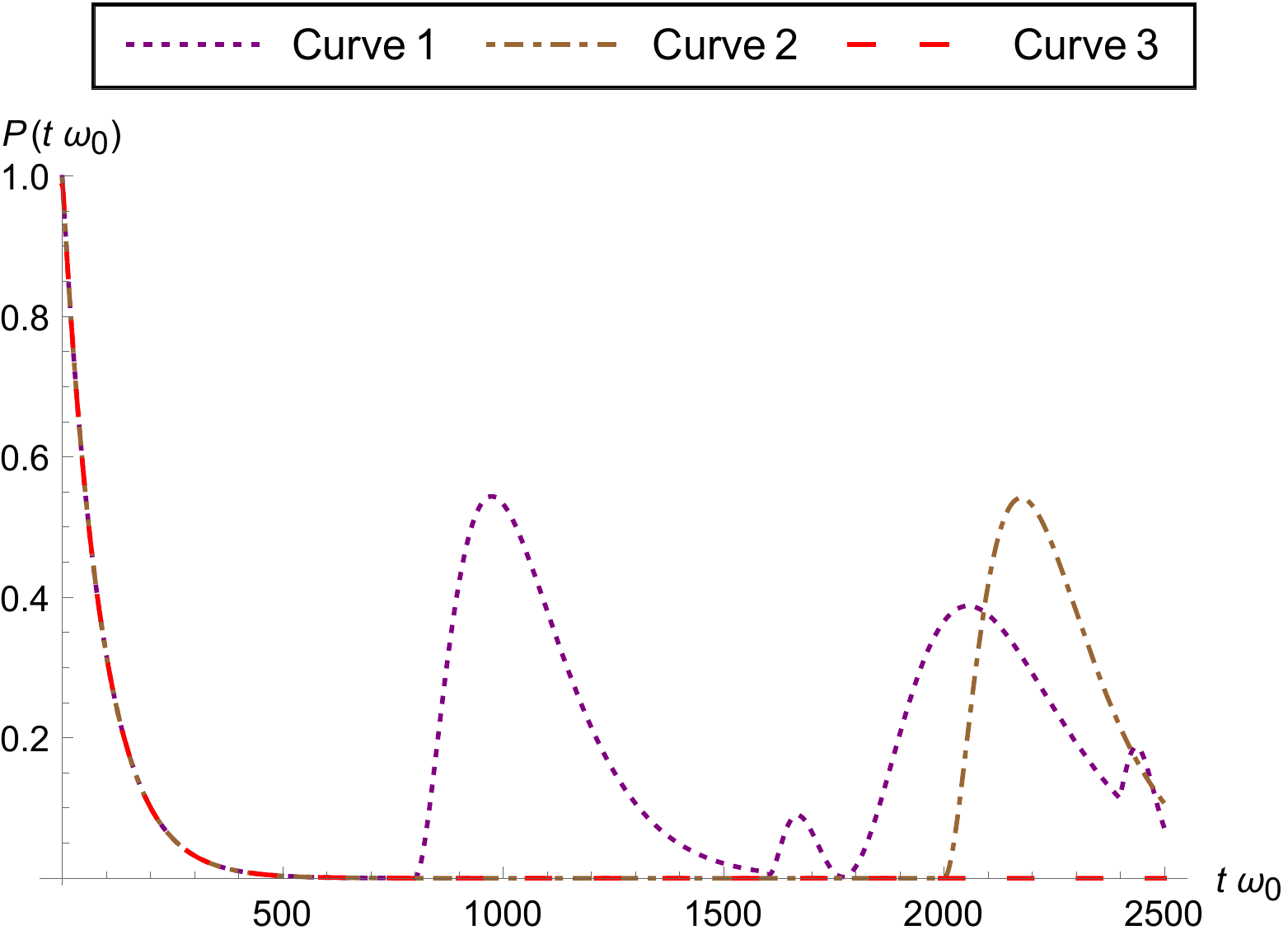}
 \caption{(Color online) Curve 1: $R\omega_0=400$, $g=\omega_0/274$;
 Curve 2:$\omega_0 R=1000$, $g=\omega_0/274$; Curve 3: $R\rightarrow\infty$, $g=\omega_0/274$. }
\label{fig6} 
\end{figure} 
\begin{figure}[b!]
\includegraphics[scale=0.4]{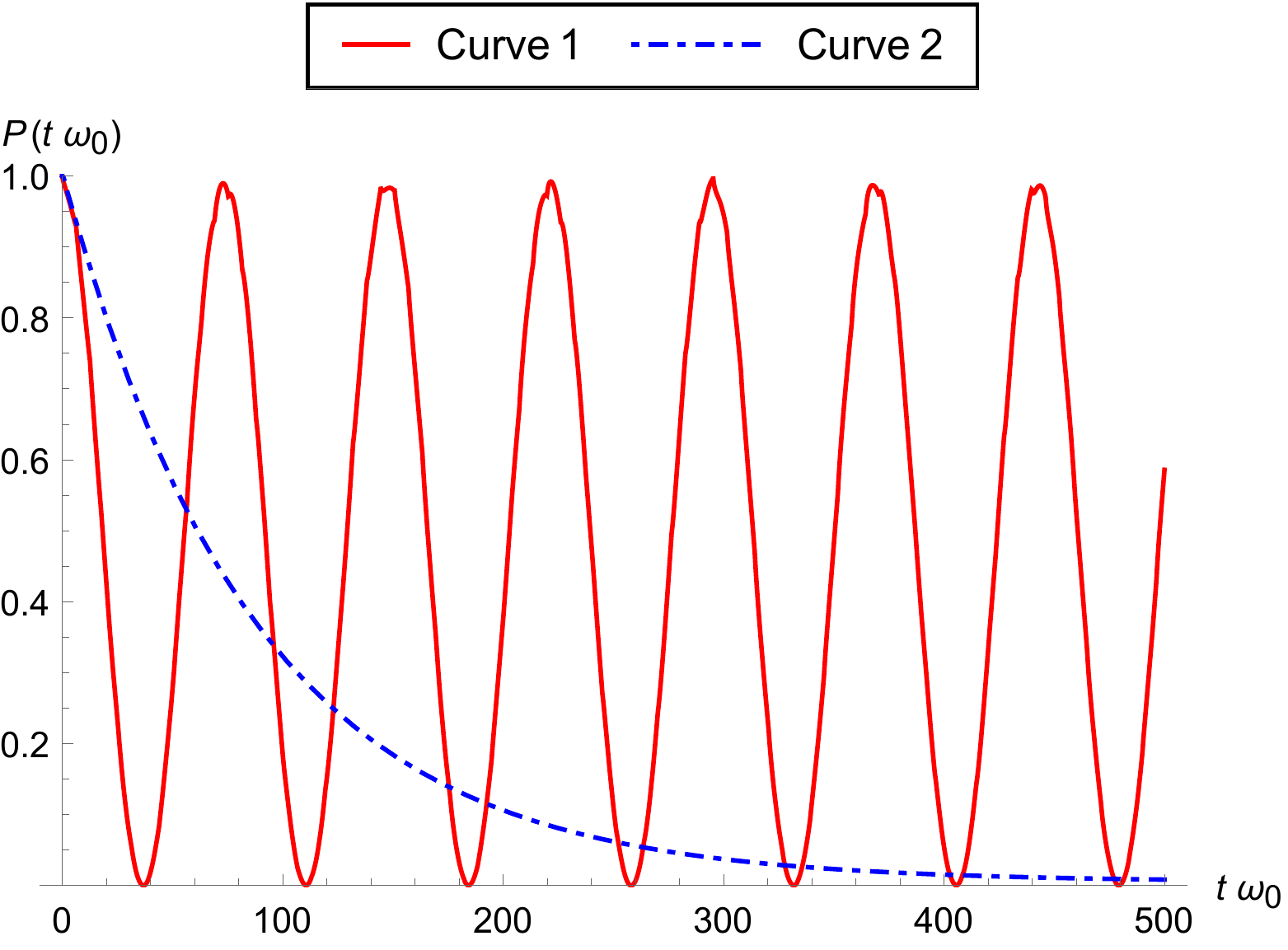}
 \caption{(Color online) Curve 1: $R\omega_0=\pi$, $g=\omega_0/274$;
 Curve 2: $R\to \infty$, $g=\omega_0/274$. }
\label{fig7} 
\end{figure} 

In Fig. \ref{fig5}, for $g=\omega_0/274$ and $gR=400/274$  it appears that $P(t)$ decreases
in time for all $t$, however, considering sufficiently large time
values, we can see in  Fig. \ref{fig6}, that $P(t)$ increases from a given time value and afterward
decreases again. In the
same figure we also depict the case in  which $gR=1000/274$, with a similar behaviour. From this results we note that
although $P(t)$ increases from a given value of time, it remains
practically at zero value for a large time interval before the first
oscillation and such time interval increases with $R$. Also, in the time interval before $P(t)$ increases, this
remains the same in both cases and practically is the same as in  the $R\to\infty$ case. In this way we have a clear
picture about how the time behaviour for $P(t)$ go from the oscillating behaviour, for $R$ finite, to the
almost exponential decay  in  free-space, as $R\to\infty$, the period of oscillation go to infinity.

\begin{figure}[b!]
\includegraphics[scale=0.4]{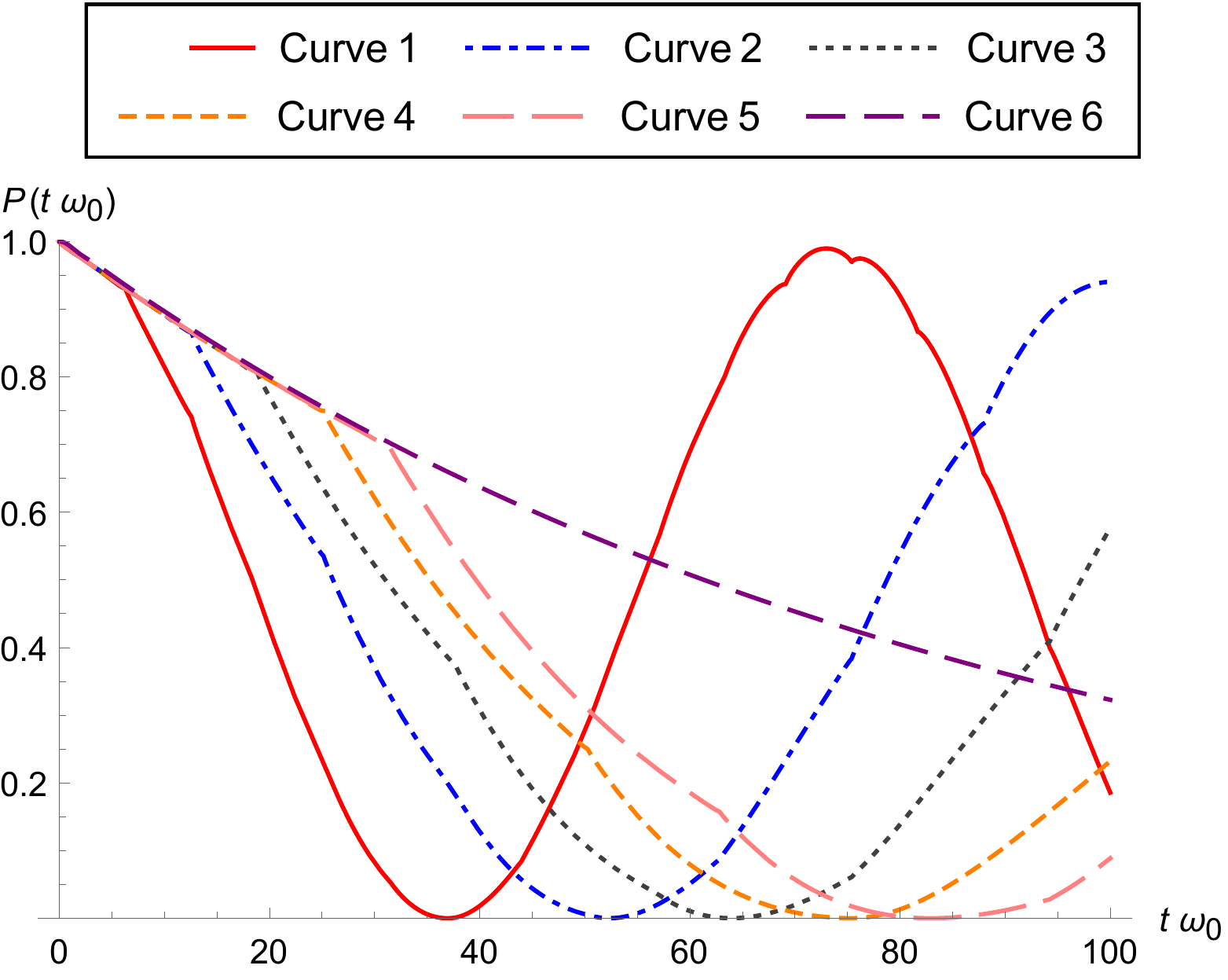}
 \caption{(Color online) Curve 1: $R\omega_0=\pi$, $g=\omega_0/274$;
 Curve 2: $R\omega_0=2\pi$, $g=\omega_0/274$;
Curve 3: $R\omega_0=3\pi$, $g=\omega_0/274$;
Curve 4: $R\omega_0=4\pi$, $g=\omega_0/274$;
Curve 5: $R\omega_0=5\pi$, $g=\omega_0/274$
 Curve 6: $R\to \infty$, $g=\omega_0/274$. }
\label{fig8} 
\end{figure} 
Although in general the spontaneous decay increases when $R$ is
increased and vice-versa, there are however
some values of $R$ for which this behaviour could be different. 
Consider the case in which $R$ takes
a value in which the cavity is in resonance with the frequency of
the atom, {\it i.e},  $\omega_0=\omega_k=\pi k/R$, $k=1,2,..$.
To be specific we consider $g=\omega_0/274$ and the minimum resonant value for $R$,  $R\omega_0=\pi$. 
We obtain the result showed in  Fig. \ref{fig7} where we also depicted the behaviour of
the $R\to\infty$ case for comparison. We note from Fig. \ref{fig7} that although in this case, the probability of the atom
to remain in its first excited level is an oscillatory function of time (we have  Rabi oscillations)
it decays more rapidly than  the free space case for initial time values. Therefore, for that resonant cavity   we have an 
enhancement of the spontaneous decay, which is more significant, for early times, before the first Rabi oscillation.
If we consider other values for $R$ resonant, we get the same conclusion, but the effect is more
appreciably in the case we just considered, $R\omega_0=\pi$, as one can
conclude from Fig. \ref{fig8}, where we display the $P(t)$ behaviour for  other resonant values of $R$. 

\section{Conclusions}

In this paper we considered the dependence of the spontaneous decay of an atom, roughly approximated by a dressed  harmonic oscillator,
on the cavity size in which it is enclosed. 
For  small cavities, we obtained analytical expressions and for cavities of arbitrary size, we carried out
numerical computations that we found in good agreement with the analytical results for sufficiently small cavities and for free-space. In general, when the cavity size  increases, the probability of spontaneous decay of the atom increases and vice versa. We obtained the well know experimental result, that for
sufficiently small cavities the probability of spontaneous decay is greatly suppressed in relation to the free case, 
whereas for large values of the cavity radius it approaches the free-space case, $R\to\infty$. On the other
hand, we found that there are some values for the cavity radius for which the spontaneous decay is increased in relation to the free case. This occurs when
$R=n\pi/\omega_0$, $n=1\ ,2\ ,3\ \ldots$, the maximum enhancement of the spontaneous decay being achieved for $n=1$.  

From the obtained results, it is not difficult to see that in the initial times,  the atom decays as if it were in free space for a time interval that increases with 
the cavity radius. This behaviour can be explained in terms of the time  the field quanta takes to go up the cavity wall and back up to the atom. Before
the field quanta comes back to the atom, it does not "know" that it is confined, therefore the spontaneous decay evolves as in the free case for a time
interval of the order $\Delta t\approx 2R/c$. For example, considering the case in which $R\omega_0=400$, we get 
 $\Delta t \omega_0=800$ (in $c=1$ units), and for $R\omega_0=1000$ we have $\Delta t\omega_0=2000$, both values
 in good agreement with our numerical computations depicted in curves 1 and 2 of Fig. \ref{fig6}. 

Finally, we would like to call attention about the dependence of the spontaneous decay on the coupling constant. Since the 
dimensionless parameter in our model is $gR$, if we fix $R$, all our conclusions remains the same in terms of the coupling constant.

{\bf Acknowledgements}\\
This work was partially supported by Brazilian agencies CNPq and CAPES.



\end{document}